\documentstyle[12pt,aaspp4]{article}

\begin{document}

\title{
Constraining the Location of Microlensing Objects by \\using the Finite Source Effect in EAGLE events\\ }

\author{Takahiro Sumi}
\affil{Solar-Terrestrial Environment Laboratory,  Nagoya University,
Nagoya 464-8601, Japan; e-mail:tsumi@stelab.nagoya-u.ac.jp }

\and

\author{Mareki Honma\altaffilmark{1,2}}
\affil{1 VERA Project Office, National Astronomical Observatory, 2-21-1 Osawa, Mitaka, Tokyo 181-8588, Japan; e-mail:honmamr@cc.nao.ac.jp }
\affil{2 Mizusawa Astrogeodynamics Observatory, NAOJ, Mizusawa, Iwate 023-0861, Japan}

\begin{abstract}
We propose a new method to constrain the location of microlensing objects 
using EAGLE (Extremely Amplified Gravitational LEnsing) events.
We have estimated the rate of EAGLE events by taking the finite-source effect into account. We found that the EAGLE event rate for using a 1-m class telescope whose limiting magnitude is $V \sim 21$ is the same as or higher than that of the ordinary microlensing events which have been found to date.
We have also found that the fraction of transit EAGLE events is large enough to detect: between $4 \sim 80 \% $ depending on the lens location.
Since the lens proper motion can be measured for a transit event, one can 
distinguish whether the lens is a MACHO (MAssive Compact Halo Object) in our halo or one of the known stars in the Large Magellanic Cloud (LMC) from the proper motion measurement for each transit EAGLE event.
Moreover, we show that the fraction of transit EAGLEs in all EAGLE events significantly depends on the lensing locations: the transit EAGLE fraction for the self-lensing case is $2 \sim 15$ times larger than that for halo MACHOs.
Thus, one can constrain the location of lens objects by the statistics of the transit events fraction.
We show that we can reasonably expect $0 \sim 6$ transit events out of 21 EAGLE events in 3 years. We can also constrain the lens population properties at a greater than $99 \%$  confidence level depending on the number of transit events detected.  We also present the duration of EAGLE events, and show how an hourly observational mode  is more suitable for an EAGLE event search program.
\end{abstract}


\section{Introduction}

Several groups have been engaged in gravitational microlensing observations 
toward the Large Magellanic Cloud (LMC) and the Small Magellanic Cloud (SMC) 
in order to investigate the nature of MAssive Compact Halo Objects (MACHOs) 
in the Galaxy's halo.
Until now, 8 microlensing event candidates have been found toward the
LMC (\cite{alc97}), and 2 candidates, including one possible binary lens
event, have been found toward the SMC (\cite{afo98}; \cite{afo99}; \cite{alb99}; \cite{alc99a}). 

The microlensing light curve informs us only of the event duration $\hat{t}$,
in which the mass, velocity and distance of the lensing object are
degenerate, and thus it is difficult to determine the mass or the distance
of the lensing object for each microlensing event.
Therefore, the nature of MACHOs in the Galactic halo still remains unclear, and one cannot 
rule out the possibility that the microlensing events are not 
caused by MACHOs in the
Galaxy's halo but by unknown populations lying between the Galaxy and the
Clouds (e.g., Zaritsky \& Lin 1997; Zhao 1998), or by normal stars in the
Clouds themselves (e.g.,Sahu 1994).

In order to understand the nature of lensing objects, some additional
information that partially or fully break the three-fold degeneracy is
required.
Over the past years several investigations have been made to extract
additional information from special types of microlensing events. Examples of which are binary events and parallax events, for which the lens proper motion and lens distance can be constrained (e.g., Schneider \& Weiss 1984; \cite{alc95}; Miyamoto \& Yoshii 1995).
A transit event, in which the lens transits the extended surface of the
source star, is another example of such types of microlensing events 
(Gould 1992, 1994; \cite{nem94}; \cite{wit94}; \cite{pen97}).
When the lens passes over the surface of the source star, the point source
approximation fails and the peak of light curve deviates from that for a
point source.
From such a light curve, one can derive the radius of the source star
scaled by the Einstein ring radius projected onto the source plane.
Since the radius of the source star can be estimated from its spectrum, one
can obtain the proper motion of the lens for transit events.
The proper motion of the lens will be of great use in investigating the
lens location, as the expected proper motion distribution of objects in the
Galaxy's halo differs significantly from that in the Clouds.
Unfortunately however, the probability for transit events is expected to be
extremely small ( $\sim 0.1\% $) for normal microlensing events (\cite{gou94}).

The probability for transit events is relatively high for
EAGLE  events, in which an
invisible star is amplified above the detection limit, because EAGLE events are 
guaranteed to have a small
impact parameter (\cite{wan96}; \cite{gou97}; \cite{nak98}).
Since there are many more faint stars below the detection limit than visible
stars that are usually used for microlensing search, the EAGLE event rate
is expected to be fairly high (\cite{nak98}). However EAGLE
events have rarely been seen in the current observation programs 
in which the brightness of visible stars are measured based on
a DoPHOT-type analysis.
Recently however, a new CCD photometry method called {`}image subtraction{'} has
been developed  (\cite{ala98}, 2000; \cite{alc99b}; I.A. Bond 1999, private communication).
This method, which compares an exposure frame directly with a reference
frame, is much more powerful for detecting EAGLE events than the current
monitoring of visible stars, and one can expect that plenty of EAGLE
events will be detected with this method in the near future.
If many EAGLE events are detected, one may be able to find several transit
events because EAGLE events have small impact parameters.
Therefore, a search for EAGLE events will possibly provide a new opportunity to 
measure the lens proper motion in transit events, and thereby to constrain strongly the lens location.
Furthermore, statistics of transit EAGLE events alone may be useful for
studying the nature of lensing objects.
The fraction of transit events depends on the ratio of the physical radius of 
the source star $R_*$ to the Einstein radius $r_E$ in source plane, i.e.
$R_* / r_E$.
In the case where the lenses are the stars in the LMC itself (self-lensing), the Einstein radius $r_E$ in source plane is much smaller than in the case where the 
lenses are MACHOs in the Galactic halo. 
Thus the fraction of transit EAGLE events out of all EAGLE events for 
self-lensing in the LMC will be much larger than that for lensing due to MACHOs
in the halo. 
Therefore the fraction of transit EAGLE events may be used to constrain
the location and the nature of lens. 

EAGLE events will thus become one of the most powerful tools for investigating the nature of lensing objects.
For these reasons, in this paper we investigate the properties of EAGLE
events, and present a detailed calculation of the probability for transit EAGLE events.
In section $\ref{sec:basic}$ we briefly summarize the basic equations, and
in section $\ref{sec:event}$ we estimate the event rate of EAGLEs.
In section $\ref{sec:ratio}$ we calculate the fraction of transit events,
and in section $\ref{sec:duration}$ we derive the event duration of EAGLEs.
Discussion on the observation strategy for EAGLE search is made in section
$\ref{sec:disc}$.

\section{Introductory description of EAGLE events}
\label{sec:basic}

\subsection{Basic equations for EAGLE events}

The amplification A of the microlensing event for 
 point source approximation is given by
\begin{equation}
    A(u) = \frac{u^2+2}{u\sqrt{u^2+4}},
\label{eq:amp}
\end{equation}
where
\begin{equation}
    u=\frac{d}{R_E}=\frac{D_s}{D_d} \frac{d}{r_E}, \quad
     R_E = \sqrt{\frac{4GM}{c^2}\frac{(D_s-D_d)D_d}{D_s}}.
\label{eq:imp}
\end{equation}
Here $D_s$ and $D_d$  are the distances from the observer to the source star, and to the lens object respectively. $d$ is the distance from the lens to the source projected onto the lens plane. $R_E$ and $r_E= (D_s/D_d)R_E$ are the Einstein radius and the projected Einstein radius in the source plane, respectively. 
The parameter $u$ denotes the impact parameter in the lens plane scaled by $R_E$.

In an EAGLE event, an invisible star with magnitude $V$ is amplified beyond the observational detection limit $V_{obs}$.
More precisely, to identify an EAGLE event, the source must become brighter than the EAGLE detection threshold $V_{\rm th}$, which is slightly brighter than the observational detection limit $V_{obs}$.
Thus, for an EAGLE event, the amplification $A$ must be larger than the threshold amplitude $A_T$, which is written as:
\begin{equation}
A_T = 10^{0.4(V-V_{\rm th})}
\end{equation}
We define the threshold impact parameter $u_T$ as the largest impact parameter for an EAGLE event with a $V_{th}$ magnitude source star.
The value of $u_T$ is determined so that $A(u_T)=A_T(V,V_{\rm th})$.
In the case of a point source and a small impact parameter, we can make the approximation that $A\approx 1/u$.  The threshold impact parameter can be then be written as (\cite{nak98}):
\begin{equation}
    u_T=10^{-0.4(V-V_{\rm th})}
\end{equation}
However, in most of the cases we are interested in, the effect of a finite source is not negligible, and this approximation is not applicable. We thus calculate the threshold impact parameter $u_T$ by taking the finite-source effect into account.

When the source size is finite, the amplification $A$ cannot be given by equation (\ref{eq:amp}), but approximated by the following expression (\cite{gou94}),

\begin{equation}
A_{\rm fin}(u,u_*) = A(u)B(z), \quad  z\equiv \frac{u}{u_*}, \quad 
       u_* \equiv \frac{D_d}{D_s}\frac{R_*}{R_E} = \frac{R_*}{r_E}
\label{eq:afin}
\end{equation}
Here $u_*$ is the source radius scaled by $r_E$ ($R_*$ is the physical radius of the source), and $z$ is
the impact parameter in the source plane scaled by the radius of
the source star $R_*$.
The function B(z) describes the ratio of the amplification of a finite source to the amplification of a point source, which is written as
\begin{equation}
B(z) = \frac{z^2}{\pi} \int_{0}^{2\pi}d\theta \int_{0}^{1/z}dq\;
         q(1 + q^2 + 2q \cos\theta)^{-1/2}. 
\label{eq:b}
\end{equation}
We show the relation of the amplification and other parameters in 
figure $\ref{fig:amp}$.

To compute the amplification using equations (\ref{eq:afin}) and (\ref{eq:b}), one has to know the physical size of the source star that has a magnitude of $V$.
In this paper, we approximate the relation between $R_{*}$ and $V$ for a main sequence star (\cite{all73}; \cite{bla63}) as follows,
\begin{eqnarray}
   R_{*}(V) &=& 10^{-0.053V+1.22}, \qquad (21 \leq V \leq 31) \label{eq:radius1} \\
            &=& 10^{-0.125V+2.78}, \qquad (15 \leq V \leq 21).
\label{eq:radius2}
\end{eqnarray}
Using equations (\ref{eq:afin}), (\ref{eq:b}), (\ref{eq:radius1}) and (\ref{eq:radius2}),
one can derive the threshold impact parameter $u_T$ from the condition that $A_{\rm th}=A_{\rm fin}(u_T,u_*)$.

\subsection{Mass and luminosity function of source stars}

The EAGLE event rate depends heavily on 
the mass function and the luminosity function of source stars in the LMC 
and the SMC. 
Following Nakamura \& Nishi (1998), we assume that a stellar initial 
mass function is in the power law form with index $\alpha$,  
that the star formation rates in the LMC have been constant for 5 
Gyr and was zero before 5 Gyr ago.
Assuming that the lifetime of the Sun is 10 Gyr, we obtain the 
present day mass function as
\begin{eqnarray}
   \phi(M) &=& CM^{-\alpha}, \hspace{70pt}  (M_l < M < 1.4M_\odot)
\label{eq:imf1} \\
        &=& 2(\frac{M}{M_\odot})^{-2}CM^{-\alpha}, \qquad  (1.4M_\odot < M < M_u)
\label{eq:imf2}
\end{eqnarray}
where $M_l$ and $M_u$ are the lower and the upper mass limit of the 
stars, respectively.
The initial mass function of the stars in the LMC are not so different from
the Salpeter IMF $\alpha = 2.35$ (\cite{hol96}).
However for field stars the slope may be as steep as $\alpha\sim 5.0$
(\cite{wes97}). 
In the following analysis we use the 
two extreme values of $\alpha$, 2.35 and 5.0.
We note that the results in the following sections do not depend strongly on the shape of the mass function beyond $M=1.4M_\odot$ because the contribution of low mass stars is much more significant than that of massive stars.

The mass-luminosity relation of the main sequence stars is expressed by
(\cite{kip90})
\begin{equation}
L(M) =L_\odot (\frac{M}{M_\odot})^3
\label{eq:masslumi}
\end{equation}

The Bolometric Correction ($B.C.$) can be approximated as
(\cite{sch71}; \cite{dav70}; \cite{joh64})   
\begin{eqnarray}
   B.C. &=& 0.6V - 11.6, \qquad (13 \leq V \leq 19) \label{eq:bc1}\\
        &=& 0, \hspace{78pt} (19 \leq V \leq 24) \label{eq:bc2}\\
        &=& -0.37V + 8.9, \hspace{16pt} (24 \leq V \leq 31)
\label{eq:bc3}                                               
\end{eqnarray}

Then we can drive the $V$ band luminosity function of stars in the LMC using
equations (\ref{eq:imf1}) to (\ref{eq:bc3}).
\begin{eqnarray}
   \phi_L(V) &=& C'10^{-(\alpha+2)(V+ B.C.)/(-7.5)}, \qquad  (15 < V < 21) \\
          &=& C''10^{-\alpha (V + B.C.)/(-7.5)}, \hspace{40pt}  (21 < V < 31) 
\label{eq:lf}
\end{eqnarray}
In this paper we will use these mass and luminosity functions.

\section{EAGLE event rate}
\label{sec:event}

In this section we estimate the EAGLE event rate.
Instead of the absolute event rate, we estimate the relative event rate normalized with the rate for normal microlensing events.
When a lens object passes the line of sight with a normalized 
impact parameter $u=d/R_E$, the source star is amplified according to equation 
(\ref{eq:amp}), where $d$ is the distance from the lens to the source
in the lens plane.
A microlensing event is defined as a phenomenon where the source star is 
amplified more than the given threshold amplification $A_{th}$.
This occurs whenever a lens enters the microlensing ``tube'' whose radius is
$u_{th}R_E$, where $u_{th}$ is the detection threshold impact 
parameter corresponding to $A(u=u_{th})=A_{th}$ (Griest 1991).
This means that all detected microlensing events have to satisfy 
$d \leq u_{th}R_E$.
The probability that a lens passes this region is proportional to the  length
of this radius $u_{th}R_E$.
For a normal microlensing event $u_{th}$ is set to be unity corresponding to 
$A_{th}=1.34$, and for an EAGLE event we use $u_T$ defined in section 
\ref{sec:basic} as $u_{th}$.
Hence the event rate is proportional to $u_{th}R_E = R_E$ for a normal 
microlensing event and $u_{th}R_E = u_T R_E$ for an EAGLE event, i.e., these event
rates are proportional to the radius of each appropriate circle in figure \ref{fig:amp} and Einstein radius $R_E$.

The event rate is proportional to the density distribution of the lens $n_s(M,D_d)$ (Griest 1991).
The event rate is also proportional to the number of source stars
which is expressed as the luminosity function of source stars $\phi_L(V)$ 
in the LMC.
Moreover, the event rate should be averaged over the probability 
distribution of the mass of the lenses using  the mass function $\phi(M)$.
Thus, the differential form of the event rate averaged over the source stars 
for normal microlensing events is written as
\begin{equation}
   d\Gamma_N \propto R_E \phi_L(V) \phi(M) n_s(M,D_d) dD_d dD_s dV dM.
   \label{eq:dgamman}
\end{equation}
Note that the integration with respect to $V$ means averaging over the source stars.
Similarly, the differential form of the rate for EAGLE events can be expressed as 
\begin{equation}
   d\Gamma_E \propto  u_T R_E \phi_L(V) \phi(M) n_s(M,D_d)dD_d dD_s dV dM
   \label{eq:dgammae}.
\end{equation}
Note that the threshold impact parameter $u_T$ appears in case of EAGLE events,
 since the event rate for EAGLE events is proportional to $u_T R_E$ instead of 
 $R_E$.
To calculate the normal and the EAGLE event rates, we must integrate them over the possible ranges of the parameters.

The normal event rate $\Gamma_N$ can be estimated as
\begin{equation}
 \Gamma_N  = C\int_{V} \int_{M} \int_{D_s} \int_{D_d} d\Gamma_N,
   \label{eq:gamman}
\end{equation}
where the integration is performed over the following ranges.
\begin{eqnarray}
D_{d,min}\le &D_d& \le D_s, \\
D_{s,min}\le &D_s& \le D_{s,max}, \\
M_l\le &M& \le M_u, \\
V_l\le &V& \le V_{obs}.
\end{eqnarray}
Here $M_l$ and $M_u$ denote the lower and the upper mass limit for the mass function, and $V_l$ denotes the luminous end of the luminosity function $\Phi_L(V)$.
Similarly, the EAGLE event rate $\Gamma_E$ can be estimated as
\begin{equation}
 \Gamma_E  = C\int_{V} \int_{M} \int_{D_s} \int_{D_d} d\Gamma_E.
   \label{eq:gammae}                                        
\end{equation}
Note that the integral ranges for $D_d$, $D_s$ and $M$ are the same as for $\Gamma_N$, but $V$ is in the range:
\begin{equation}
V_{obs}\le V \le V_u.
\end{equation}
Note that $V_u$ denotes the faint end of the luminosity function.

Since the constant C is equal in both equations, one can easily obtain the ratio $\Gamma_E / \Gamma_N$ by integrating above equations numerically.
As for the integral ranges for the luminosity in the LMC stars, we assume that $V_l = 15$ and $V_u = 31$.
We note that the results are not affected strongly by stars outside this magnitude range.
We also assume $V_{obs}=21$, which is a typical observational limit for current microlensing programs.
We consider two different values for the threshold magnitude; $V_{th}$ = 19 and 20 .

We calculate these integrals in the case where the lenses are stars in the LMC itself (self-lensing) and in the case where the lenses are MACHOs in the halo.
For these two cases, there are the following differences which arise in the integral ranges of ($M$, $D_s$ and $D_d$) and the functions of ($n_s(M,D_d)$ and $\phi(M)$).

\begin{itemize}

\item[i)]
In the case where the lenses are known stars in the LMC itself,
for simplicity, we assume a constant spatial density distribution of 
the lens $n_s(M,D_d)$ with depth $d_{max}=D_{s,max} - D_{s,min}$.
For the depth of the LMC $d_{max}$, one finds that the bar can be as thin as
the disk itself (\cite{bin87}), and the thickness of the LMC disk is about 300pc
(\cite{wes91}).
Since there is not any more accurate information,
 we adopt two values, 300pc and 1kpc. 
We also assume that the distance to the LMC is 50 kpc, and also assume $M_l = 0.1M_\odot$, $M_u=50M_\odot$ for the lens mass distribution expressed by equations
(\ref{eq:imf1}) and (\ref{eq:imf2}) in the case of self-lensing.

\item[ii)]
In the case where the lenses are MACHOs in the halo, we fix $D_s = 50$ kpc
 and integrate with $D_d$ from $D_{s,min}=0$ to $D_s = 50$ kpc.
We also adopt the standard halo model for the spatial distribution of the lens 
$n_s(M,D_d)$ as follows (Griest 1991):
\begin{equation}
   n_s(M,D_d) = \frac{\rho}{M}=\frac{\rho_0}{M}\frac{a^2+r_0^2}{a^2+r^2}
   \label{eq:halo}
\end{equation}
where $\rho_0=0.0079$ is the local mass density, $r$ is the Galactocentric radius,
$r_0 = 8.5$ kpc is the Galactocentric distance of the Sun, and $a=5$ kpc is the core radius. 
We also assume a delta function mass distribution for MACHOs, and we take two values of $M$; $M= 0.1 M_\odot$ and $1.0 M_\odot$. 
\end{itemize}

The results for $V_{\rm th}=$ 19 and 20 are listed in Table \ref{tbl:eventrate},
and the probability distribution of EAGLE events as a
function of the source magnitude $V$ are shown in figure $\ref{fig:v}$.
From these results we conclude the following:

\begin{itemize}
\item[1)]
From the results in table \ref{tbl:eventrate} we can expect the EAGLE event rate is of the same order ($\sim 7$ events/yr) as the rate for normal events when $\alpha=2.35$, and much larger ($70 \sim 170$ events/yr) 
when $\alpha=5.0$. Here we assume the normal event rate $\Gamma_N = 4$ events/yr (\cite{alc97}).
Thus, the event rate for EAGLEs is high enough to allow the events to be detected.

\item[2)]
The event rate for EAGLEs for $\alpha=5.0$ is $8 \sim 25$ times higher
than that for $\alpha=2.35$.
This strong dependence of the EAGLE event rate on $\alpha$ indicates that 
we can obtain some information on the power index $\alpha$  of the mass 
function of stars in the LMC from the EAGLE event rate.

\item[3)]
When $\alpha = 5.0$, the EAGLE event rate for halo MACHOs is 3 times 
higher than that for the self-lensing case. So if $\alpha = 5.0$, it is possible to determine
whether the lens masses are MACHOs in the halo or known stars in the LMC itself from the event rate of EAGLEs.

\item[4)]
In figure  $\ref{fig:v}$, we can see the EAGLE event probability decrease for large $V$, though the number of stars increases with increasing $V$.
The reasons  are that $u_{T}$ is small for large $V$ and the source can not 
 be amplified enough to be observed due to the finite-source effect.
Also, if $\alpha = 5.0$, the distribution of the source star magnitudes may be useful for determining whether the lens masses are MACHOs in the halo or are stars in the LMC, because this distribution is significantly different in each case (see figure $\ref{fig:v}$).
\end{itemize}

\section{Fraction of finite-source transit events}
\label{sec:ratio}

In this section, we investigate the fraction of transit EAGLE events in all EAGLE events.
The finite-source effect appears when $z$ becomes smaller than about 2
(\cite{gou94}), where $z \equiv u/u_{*}$ is the impact parameter (in the source plane) 
scaled by the radius of the source star $R_{*}$ (see section 2) and $u_*$ is 
defined in equation (\ref{eq:afin}) as the source star radius normalized by the Einstein radius in the source plane.
However, when $1<z<2$, the effect is not strong and it is quite difficult to determine whether the event is a transit event or a non-transit event (\cite{pen97}).
Therefore, here we define a transit event as an event with $z<1$.

The event rate of EAGLEs is proportional to $u_T R_E$ (see section \ref{sec:event}), and similarly, the probability that the lens transits the surface of the
source star is proportional to $u_* R_E$ .
In short, these rates are proportional to the radius of the corresponding circle in figure \ref{fig:amp}.
 
Thus, the ratio of the transit EAGLE event rate to the total EAGLE event rate is proportional to $u_*/u_T$.
In order to estimate the mean value of the fraction of transit EAGLE events, one
must integrate this ratio over all possible parameter ranges as above, in section \ref{sec:event}.
The mean value of this fraction can be estimated by evaluating
the following:
\begin{equation}
\langle \frac{u_{*}}{u_T} \rangle
     = \frac{1}{\Gamma_E }
       C\int_{M} \int_{V} \int_{D_s}
       \int_{D_d} \frac{u_{*}}{u_T}\; d\Gamma_E 
   \label{eq:ratio}
\end{equation}
The results for $V_{\rm th}= 19$ and $20$ are shown in Table \ref{tbl:ratio}.
From these results we conclude the following:

\begin{itemize}
\item[1)]
The fraction of transit events in all EAGLE events is much higher than that for normal events ($\sim 0.1\%$) (Gould 1994).
So we can expect a few transit EAGLE events per year using a 1-m class telescope
depending on the lens location. 
If we obtain an accurate light curve and a good estimate of the radius of the 
source star by follow-up observations, we can get the proper motion of the lens for each event (\cite{gou94}).
Since the lens proper motion is useful for constraining the lensing location
(and so the lens population),
we can determine the location of the lens for each event.

\item[2)]
Moreover, the fraction of transit events strongly depends on whether the lens is in the halo or in the LMC.
The transit fraction for the self-lensing case for $\alpha = 2.35$ is $4 \sim 15$ times larger than that for halo MACHO lensing case (for $\alpha = 5.0$, $2 \sim 5$ times larger).
We can constrain the power index $\alpha$ using the event rate (see Table \ref{tbl:eventrate}) and the distribution of the source star magnitude (see figure $\ref{fig:v}$). 
Hence, if we can measure the fraction of transit events in all EAGLE events, 
we will be able to constrain the lens location (and so the lens population) 
statistically.
\end{itemize}

So the EAGLE event search will be of great importance in investigating the 
location and the nature of the lensing objects.
We discuss some quantitative examples about these in section \ref{sec:disc}.

\section{Event duration}
\label{sec:duration}

In this section, we estimate the duration, $t_E$ of an EAGLE event.
We define the duration of an EAGLE event $t_E$ as the time when the source star becomes brighter than the observation apparent magnitude threshold $V_{obs}$.
The critical amplification $A_{obs}$ is written 
as $A_{obs}=10^{0.4(V-V{obs})}$. If $A_{obs} >2.5$, we use equation
(\ref{eq:afin}) to derive $u_{obs}$. On the other hand,  when 
 $A_{obs} \leq 2.5$, we use the standard equation ($\ref{eq:amp}$),
because equation ($\ref{eq:afin}$) is invalid and the finite-source effect  is negligible for this case.
The mean value of $t_E$ for the event with source magnitude 
$V$ can be written as
\begin{eqnarray}
  \langle t_E(V) \rangle &=& 2\hat{t} \frac{1}{u_T} \int_0^{u_T} 
         \sqrt{u_{obs}^2 - u_{min}^2}du_{min},\quad \hat{t} = \frac{R_E}{v_t} \\
       &\simeq& 1.94 \hat{t} u_{obs}(V), \qquad  (V_{\rm th}=20) \\
       &\simeq& 1.98 \hat{t} u_{obs}(V), \qquad  (V_{\rm th}=19)
  \label{eq:tev}
\end{eqnarray}
where $\hat{t}$ is the true duration of the microlensing event
 and $v_{t}$ is the transverse velocity of the lens object relative to the observer-source line of sight.
We assume $v_t= 220$ km/s for the MACHOs in the halo and $v_t= 30$ km/s  for the stars in the LMC.

The mean value of $\langle t_E(V) \rangle$ averaged over all possible values of distance, lens mass and source magnitude is:
\begin{equation}
   \langle t_E \rangle = \frac{1}{\Gamma_E}
         C\int_{M} \int_V \int_{D_{s}}
       \int_{D_d} \langle t_E(V) \rangle d\Gamma_E 
   \label{eq:te}
\end{equation}

The results for $V_{\rm th}= 19$ and $20$ are shown in Table \ref{tbl:te},
and the probability distribution of $t_E$ for EAGLE events is 
shown in figure $\ref{fig:te}$.
In figure $\ref{fig:te}$, 
the distributions are shifted to shorter durations for $\alpha =5.0$ with respect to those for $\alpha = 2.35$ because the number of dim sources for EAGLEs increases.
We can also see a cut-off in the probability curves at smaller $t_{E}$ for 
self-lensing.
This is because the event rate of EAGLEs whose impact parameter is 
smaller than some threshold value decreases according 
to the finite-source effect.
From figure $\ref{fig:te}$, we can see that
the duration of the EAGLE is usually short (1 day $\sim$ 40 days),
especially when $\alpha = 5.0$.
Hence, the nightly observational mode currently undertaken is not adequate for detecting EAGLE events, but an hourly observational mode is more suitable for detecting EAGLE events.
If observations are made several times per night, there will be sufficient time resolution to detect the EAGLE events and to issue an alert for an occurring EAGLE event.

\section{Discussion and Conclusion}
\label{sec:disc}

We have seen that the EAGLE event rate is as high as that for normal events.
Since the duration of an EAGLE event is usually short (1 day $\sim$ 40 days), the nightly observation program currently undertaken by most groups is not adequate. Hourly observation is necessary for finding EAGLE events.
Moreover, monitoring with a 1-m class telescope is sufficient to search for EAGLEs and to make follow up observations near the peak, but not enough to follow up the wing of the light curve. This is because the source is visible only at the peak of the light curve but invisible during most of the remainder of the event.
To get overall structure of the light curve, one has to observe the event with
a larger telescope for a longer period (but we do not need to observe so frequently).

Current microlensing experiments are being carried out toward very dense stellar fields.
So, the determination of the absolute flux of the source has been an issue in
microlensing because of source star blending.
In the case of EAGLE events, the source stars are fainter than the detection limit and thus the blending of brighter stars is significant, making the problem worse.
Attempts have been made to correct the blending effects for individual events by
introducing an additional parameter, the blended flux, but this method suffers from
very large uncertainties in the derived lensing parameters due to degeneracies
among the parameters (\cite{ala97}; \cite{han97}).
One solution to correct for blending problem is to use the  Hubble Space Telescope (HST).
The high resolving power of the HST and the color information from ground-based
observations enable us to uniquely separate the lensed source star from
blended stars (\cite{han97}).
The image subtraction method (\cite{ala98}, 2000; \cite{alc99b}; I.A. Bond 1999,
private communication) should be used to locate the centroid of the lensed flux 
in the ground-based event image accurately to find the lensed star in the HST 
image. 
Other high resolution telescopes like the VLT are also good candidates for performing follow-up observations.
Using a large telescope is also required for spectroscopic observations of the 
source star, from which one can estimate the source radius.
We can do this even after the event is finished.

Using the image subtraction method has another advantage in microlensing experiments.
The color change induced by the limb-darkened extended source helps the measurement of the proper motion of the lens and increases the possible number of proper motion measurement
events (\cite{wit95}; \cite{gou96}).
Han (\cite{han00}) found the color change measurement by using the image
subtraction method enables one to obtain the same information about the lens and the
source star, but with significantly reduced uncertainties due to the absence of
the blending effect.
So measuring the color change with differential photometry helps to derive the proper motion of the lens along with the other microlensing parameters,
and at least, inform us whether a lens transited the surface of the source or
not without requiring the absolute baseline flux.
This information is useful in understanding the location
of the lens object which we discuss below.

In order to perform follow-up monitoring and spectroscopic observations, one has
to detect an EAGLE event in real-time and issue an alert.
For the real-time detection of EAGLE events, the image subtraction method
is more suitable than the DoPHOT analysis or the pixel analysis, since the image subtraction method can detect the luminosity variation at any position of the fields, even where no star was identified previously.
Thus, the most reasonable and practical observation strategy is to observe hourly (with a 1-m telescope) and to perform the real-time analysis with the
subtraction method. An alert can then be issued, to observers around the world enabling frequent observations around the event peak. After that, high resolution observations with larger telescope should be carried out to determine the baseline flux of the source.

As seen in section \ref{sec:ratio}, a large fraction of EAGLE events is likely to be transit EAGLE events.
If we can issue an alert for such an EAGLE event, observations with the global network will allow us to obtain a precise light curve near the maximum amplification.
Because the transit time is $\approx 13$ hours for self-lensing and $\approx 2$ hours for halo MACHOs, it is possible to measure precisely the light curve in the transit region.
Since the overall structure of the light curve can be determined by follow-up
observations with larger telescopes, one can obtain the source radius scaled by
Einstein radius $u_*$ from the full light curve and from the color change
measurements due to a limb-darkened extended source (\cite{han00}).
If one can obtain $u_*$ and also the radius of the source star $R_*$ from the color of the source star, one can obtain the Einstein radius $r_E$ projected in source plane.
Since the event duration $\hat{t}$ can be also determined from the full light curve, one can measure the proper motion of the lens object for a transit EAGLE event (\cite{gou94}).
The proper motion of the lens object is of great use in determining the lens location.
Since the proper motion of objects in the halo is $30 \sim 40$
times larger than that of stars in the LMC, one can know whether the lensing object exists in the halo or not from the proper motion of transit EAGLE events.
Even if we cannot derive an accurate proper motion of lens,
the fraction of transit events can be used to constrain the lens location, because that ratio for the self-lensing case is $2 \sim 15$ times larger than that for the halo MACHOs case.

To demonstrate how we can constrain the nature of microlensing objects, we show in figure 
$\ref{fig:poiss}$ the probability distribution of the transit events for both the halo lensing case and the LMC self-lensing case.
For the typical parameters $\alpha = 2.35$, $V_{th} = 20$ and a detection
efficiency of 50\%, one can expect to find $\sim 21$ EAGLE events after a 3
year observation period of 11 square degrees of the LMC central region
(as the MACHO collaboration does).
In these 21 EAGLE events, we can expect 1.1 (0.4) and 6.1 (4.6) transit events for halo MACHOs whose mass is $0.1 (1.0)M_\odot$ and for self-lensing with $d_{max}=300$ (1k) pc, respectively.
So if we do not detect any transit events in 3 years, we can reject the possibility of self-lensing at more than $99\%$ confidence level
according to Poisson statistics.
In table \ref{tbl:conf}, we show some other results on how strongly we can discriminate between the two possible locations of the lensing objects based on Poisson statistics.
It is possible that we may be able to constrain strongly the lens location based on the 3-year statistics of transit EAGLE events.

In conclusion, the study of EAGLE events is very effective for constraining the nature of lens
objects.
If an EAGLE search is made for a few years, we can form strong constraints on lens objects in microlensing events.

\acknowledgments
We are grateful to Prof. Yasushi Muraki for his supervision, and also to Yukitoshi Kan-ya, Yutaka Matsubara and Philip Yock for their helpful comments.
We would like to acknowledge the careful reading our manuscript with Nicholas 
Rattenbury.

\clearpage

\figcaption[fig1.ps]{A schematic view of the parameters for an EAGLE event 
with and without the finite source effect. The thin and dot-dashed lines
are the visible and invisible light curves for a point source. The thick curve
is the visible light curve from a finite source. The outer thin circle
is the Einstein radius, and the other impact parameters defined in
section $\ref{sec:basic}$ are also shown.
When $u$ becomes smaller than $u_{obs}$, the star becomes visible, 
and when $u$ becomes smaller that $u_T$, the event is identified 
as an EAGLE event. \label{fig:amp}}

\figcaption[fig2.ps]{The probability of an EAGLE event as a function of the source magnitude $V$ for $V_{th} = 19$. The upper and lower panels are for $\alpha=$ 2.35 and 5.0 respectively.
The bold  solid and dashed lines correspond to self-lensing for 
$d_{max} = 300pc$ and $1kpc$ respectively.
The thin solid and dashed lines correspond to halo MACHOs with masses of $M = 0.1M_{\odot}, 1.0 M_{\odot}$ respectively.
Although the number of the stars with magnitude $V$ increases as $V$ increases, the EAGLE event probability decreases for large $V$ because of small $u_{T}$ values and the finite-source effect.
When $\alpha = 5.0$, the distribution of $V$ is significantly different
for the case of self-lensing and the halo MACHOs. \label{fig:v}}

\figcaption[fig3.ps]{The probability distribution of the EAGLE event duration
$t_E$ for $V_{th}=19$. The upper and lower panels are for
$\alpha = 2.35$ and $5.0$ respectively.
The bold solid and dashed curves correspond to LMC self-lensing with 
$d_{max} = 300pc$ and $1kpc$ respectively.
The thin solid and dashed curves correspond to halo MACHOs with
MACHO masses of $M = 0.1M_{\odot}$ and  $1.0 M_{\odot}$ respectively.
The distributions are shifted to smaller $t_{E}$ values for $\alpha =5.0$ than 
that for $\alpha = 2.35$ because the number of dim sources for EAGLEs increases.
We can see a cut-off in the low $t_{E}$ side of the curves in the  self-lensing case.
This is because the event rate of EAGLEs whose impact parameter is
smaller than some threshold value decreases due 
to the finite-source effect.
\label{fig:te}}

\figcaption[fig4.ps]{The probability distribution of the number of transit 
EAGLE events out of 21 EAGLE events in 3 years according to Poisson statistics 
for $\alpha = 2.35$, $V_{th}=20$ and a detection efficiency of 50\%.
The bold solid and dashed curves correspond to self-lensing for
$d_{max} = 300pc$ and $1kpc$ respectively.
The thin solid and dashed curves correspond to halo MACHOs with 
MACHO masses $M = 0.1M_{\odot}$ and $1.0 M_{\odot}$ respectively.
One can distinguish the lens population using the number of transit events.
\label{fig:poiss}}

\begin{deluxetable}{crrrrrrrrrrr}
\footnotesize
\tablecaption{The ratio of EAGLE event rate to normal event rate
 \label{tbl:eventrate}}
\tablewidth{0pt}
\tablehead{
\colhead{$V_{th}$} & \colhead{$\alpha$} & \colhead{halo}   & \colhead{halo}   &
\colhead{self}  & \colhead{self} \\
\colhead{(mag)} & \colhead{} & \colhead{($M=0.1M_{\odot}$)}
& \colhead{($M=1M_{\odot}$)}   &
\colhead{($d_{max}=300pc$)}  & \colhead{($d_{max}=1kpc$)} 
}
\startdata
19 & 2.35 & 0.72 & 0.73 & 0.60  & 0.65\\
19 & 5.0  & 15.7 & 17.9 & 4.71  & 6.46\\
20 & 2.35 & 1.83 & 1.83 & 1.67  & 1.73\\
20 & 5.0  & 44.5 & 45.0 & 18.6  & 23.2\\
 \enddata

\end{deluxetable}

\begin{deluxetable}{crrrrrrrrrrr}
\footnotesize
\tablecaption{The fraction of transit events in all EAGLEs \label{tbl:ratio}}
\tablewidth{0pt}
\tablehead{
\colhead{$V_{th}$} & \colhead{$\alpha$} & \colhead{halo}   & \colhead{halo}   &
\colhead{self}  & \colhead{self} \\
\colhead{(mag)} & \colhead{} & \colhead{($M=0.1M_{\odot}$)}
& \colhead{($M=1M_{\odot}$)}   &
\colhead{($d_{max}=300pc$)}  & \colhead{($d_{max}=1kpc$)} 
}
\startdata
19 & 2.35 &  0.09 &  0.04  &  0.45  & 0.34\\
19 & 5.0  &  0.40 &  0.25  &  0.82  & 0.72\\
20 & 2.35 &  0.05 &  0.02  &  0.29  & 0.22\\
20 & 5.0  &  0.29 &  0.12  &  0.67  & 0.59\\
\enddata

\end{deluxetable}

\begin{deluxetable}{crrrrrrrrrrr}
\footnotesize
\tablecaption{The mean EAGLE event duration $\langle t_E \rangle$
 \label{tbl:te}}
\tablewidth{0pt}
\tablehead{
\colhead{$V_{th}$} & \colhead{$\alpha$} & \colhead{halo}   & \colhead{halo}   &
\colhead{self}  & \colhead{self} \\
\colhead{(mag)} & \colhead{} & \colhead{($M=0.1M_{\odot}$)}
& \colhead{($M=1M_{\odot}$)}   &
\colhead{($d_{max}=300pc$)}  & \colhead{($d_{max}=1kpc$)}
}
\startdata
19 & 2.35 &  21.2 &  66.6 &  25.8 & 43.8  \\
19 & 5.0  &   3.6 &  10.0 &   7.7 & 10.6  \\
20 & 2.35 &  20.6 &  65.2 &  22.9 & 40.3  \\
20 & 5.0  &   3.1 &   9.8 &   5.0 &  7.4  \\
\enddata

\tablecomments{$\langle t_E \rangle$ is given in days.  }
\end{deluxetable}

\begin{deluxetable}{crrrrrrrrrrr}
\footnotesize
\tablecaption{Confidence levels for rejecting each lens mass configuration
 \label{tbl:conf}}
\tablewidth{0pt}
\tablehead{
\colhead{transit event} &  \colhead{halo}   & \colhead{halo}   &
\colhead{self}  & \colhead{self} \\
\colhead{(number)} & \colhead{($M=0.1M_{\odot}$)}
& \colhead{($M=1M_{\odot}$)}   &
\colhead{($d_{max}=300pc$)}  & \colhead{($d_{max}=1kpc$)}
}
\startdata
0 & ---  &  ---  &  99.8 &  99.0 \\
1 & ---  &  ---  &  98.4 &  94.4 \\
4 & 97.4  & 99.92   &  ---  &  ---  \\
5 & 99.5  & 99.994  &  ---  &  ---  \\
6 & 99.92 & 99.9996 &  ---  &  ---  \\
\enddata

\tablecomments{The values for confidence level of rejection are in \% 
for each number of transit events detected in 3 years for 
$\alpha = 2.35$, $V_{th} = 20$ and a detection efficiency of 50\%,  }
\end{deluxetable}

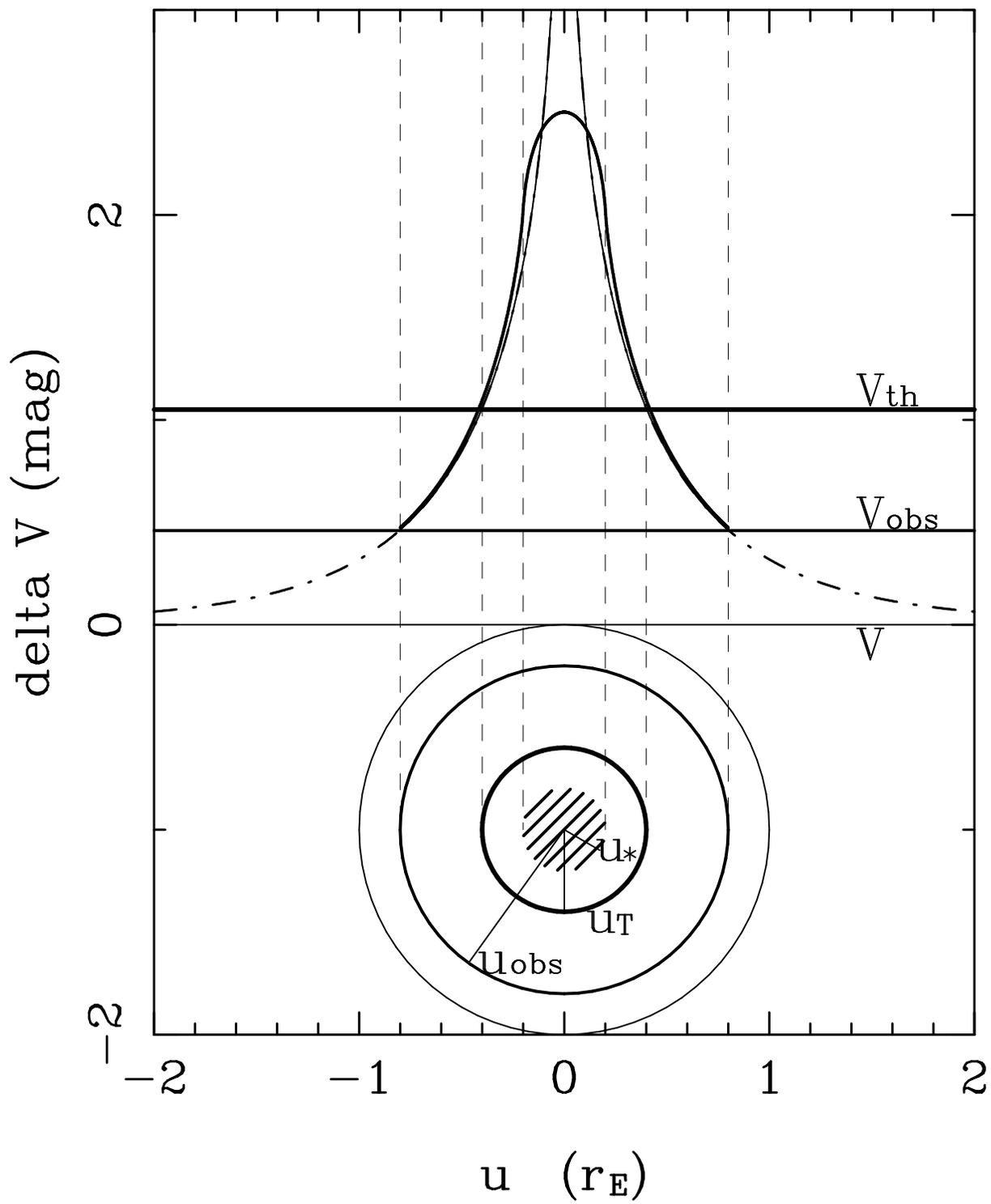
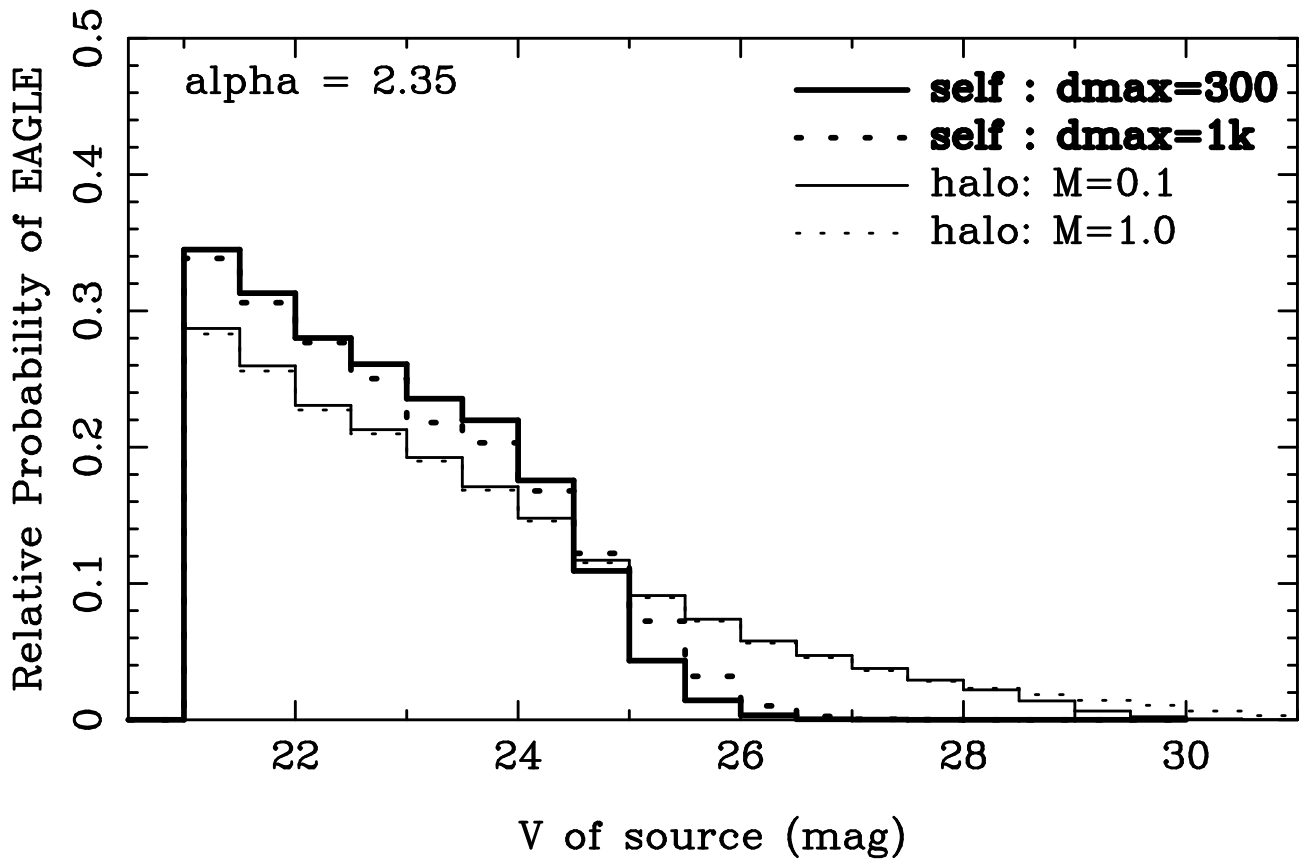
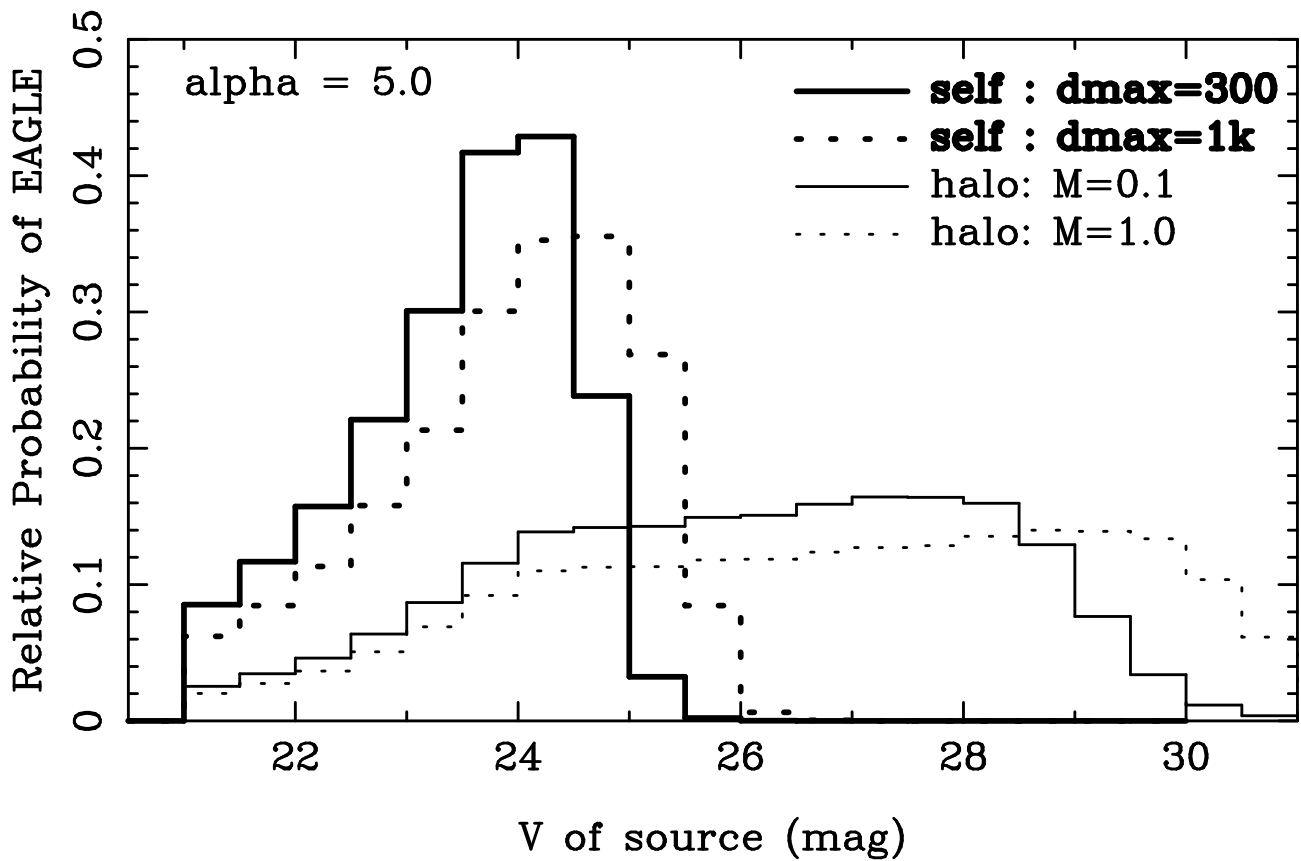
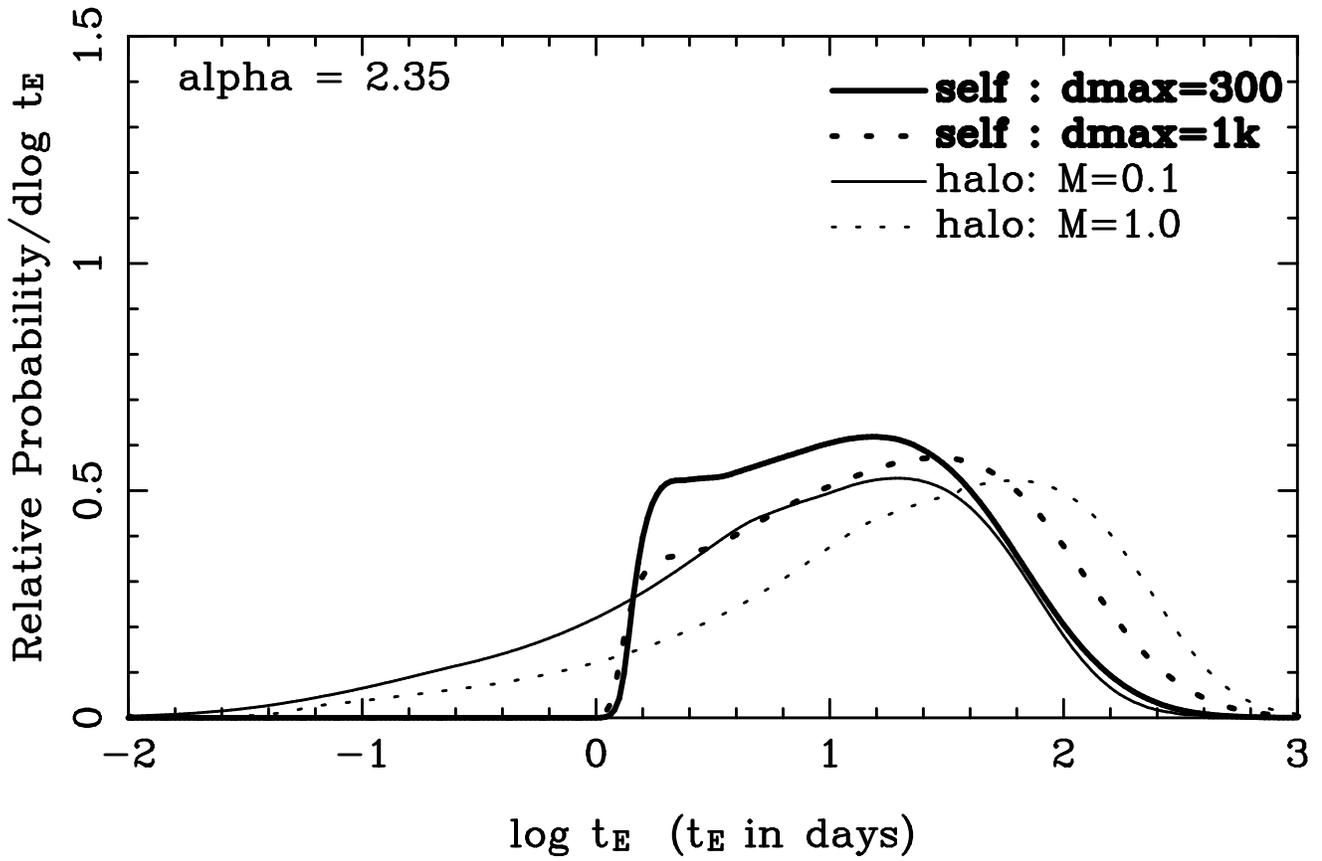
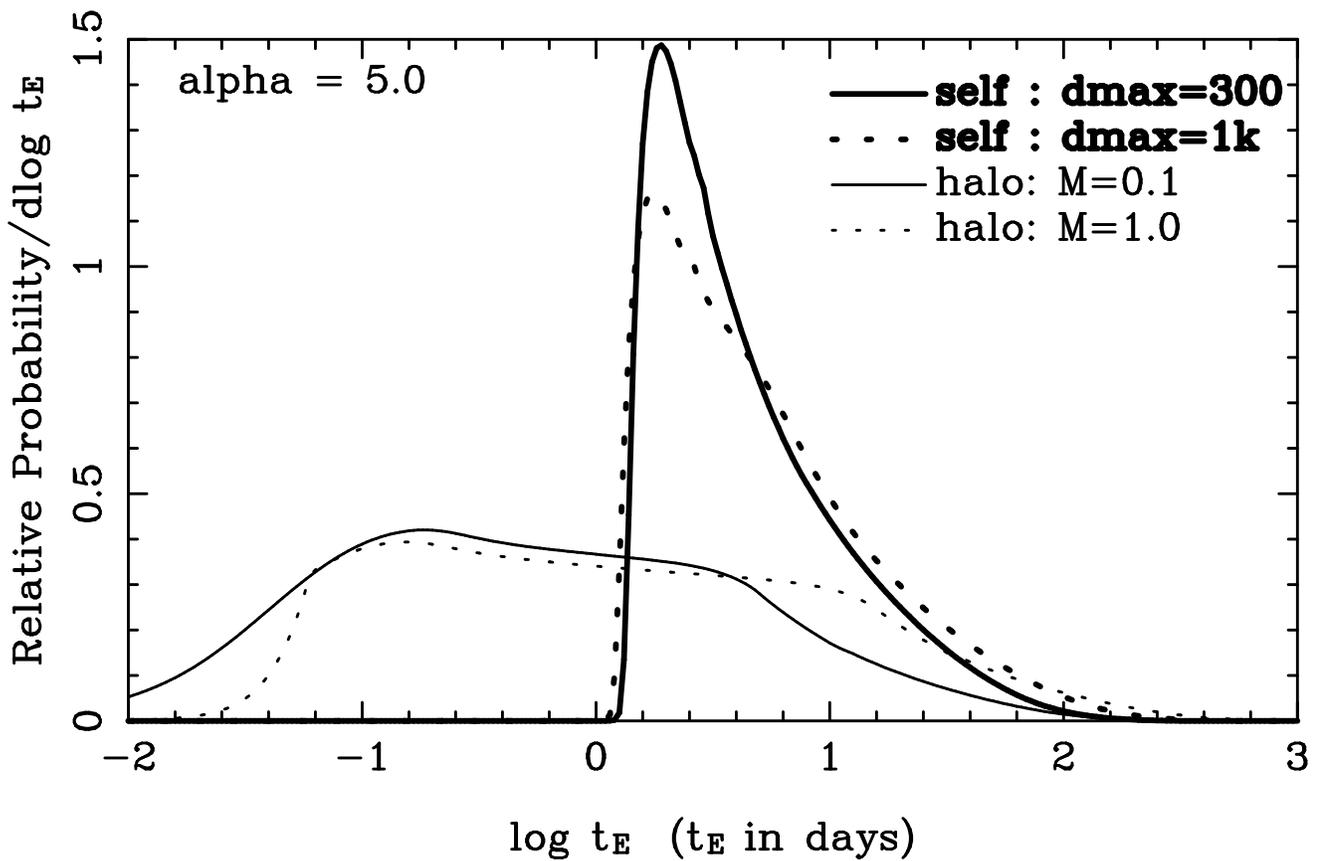
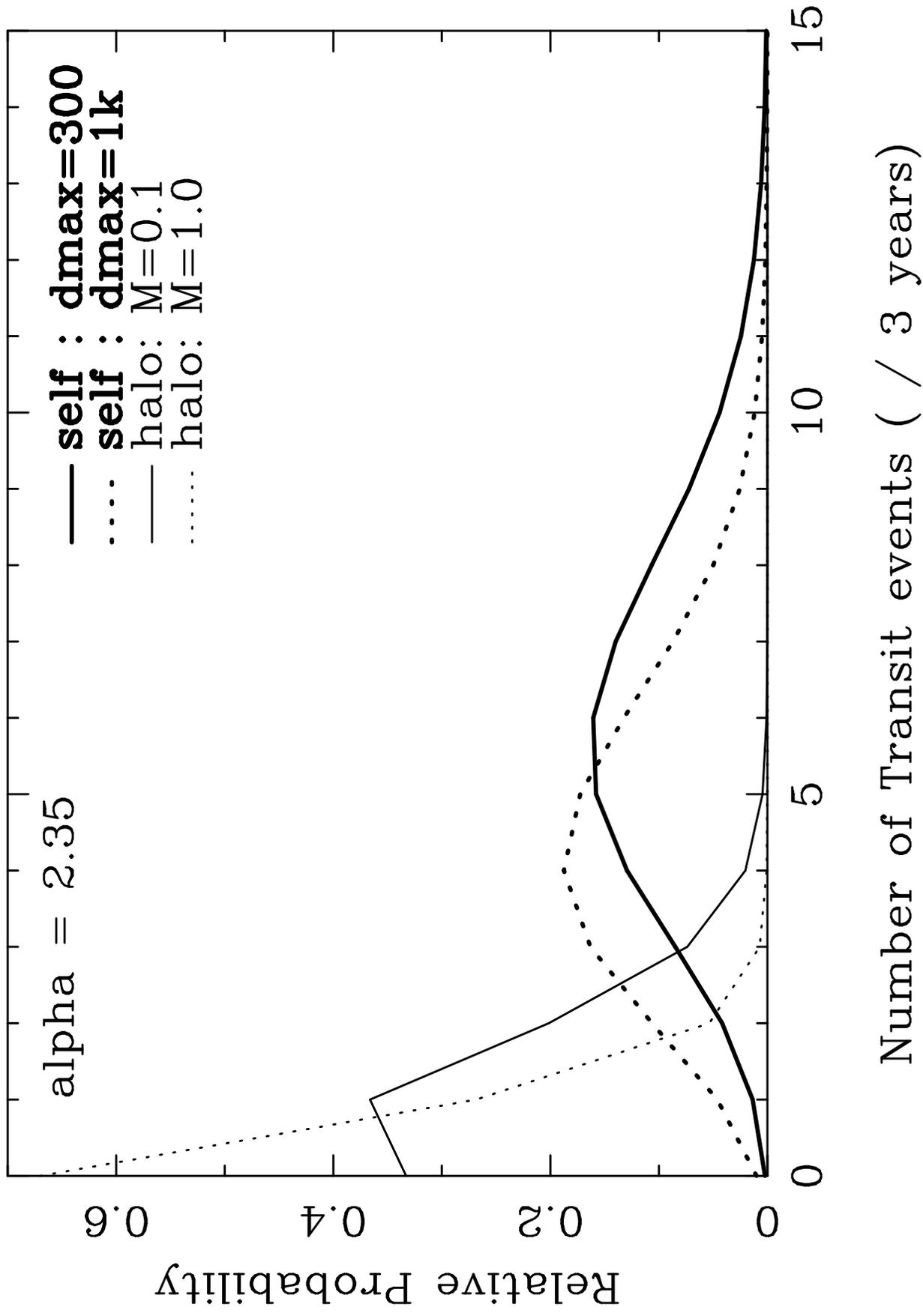
\end{document}